\documentclass{ws-ijmpcs}

\begin{document}

\markboth{Daniel M\"uller}
{Homogeneous Solutions of Quadratic Gravity}

%
\catchline{}{}{}{}{}
%

\title{HOMOGENEOUS SOLUTIONS OF QUADRATIC GRAVITY}

\author{DANIEL M\"ULLER}

\address{Instituto de F\'\i sica - Universidade de Bras\'\i lia \\
Cxp 04455, 70919-900, Bras\'\i lia, DF, Brazil \\
muller@fis.unb.br}
\maketitle

\begin{history}
\received{3 June 2011}
\end{history}

\begin{abstract}
It is believed that soon after the Planck time, Einstein's general relativity theory should be corrected to an effective quadratic theory. In this work we present the $3+1$ decomposition for the zero vorticity case for arbitrary spatially homogenous spaces. We specialize for the particular Bianchi $I$ diagonal case. The $3-$ curvature can be understood as a generalized potential, and the Bianchi $I$ case is a limiting case where this potential is negligible to the dynamics. The spirit should be analogous, in some sense to the BKL solution. In this sense, a better understanding of the Bianchi $I$ case could shed some light into the general Bianchi case.

\end{abstract}

\ccode{PACS numbers: 98.80.Cq, 98.80.Jk}

\section{Introduction}

The semi-classical theory consider the back reaction of quantum fields in a classical geometric
background. It began about forty years ago with De Witt Ref. \refcite{DeWitt}, and since then, its
consequences and applications are still under research, see for example Ref. \refcite{Hu}.

Different from the usual Einstein-Hilbert action, the one loop effective gravitational action
surmounts
to quadratic theories in curvature, see for example Refs. \refcite{DeWitt,liv}.
It is the gravitational version of the Heisenberg-Euler electromagnetism. As it is well
known, vacuum polarization introduces non linear corrections into Maxwell
electrodynamics,\cite{schwinger} first obtained by Heisenberg- Euler.\cite{heisenberg}

This quadratic gravity was previously studied by Starobinsky.\cite{S} It is of interest
for example in the context of the final stages of evaporation of black holes,
inflationary theories,\cite{coule} in the approach to the singularity,\cite{montani} and
also in a more theoretical context.\cite{cotsakis}

The effective gravity, was apparently first
investigated in Tomita's article Ref. \refcite{berkin} for general Bianchi $I$
spaces. They found that the presence of anisotropy contributes to the
formation of the singularity. Berkin's work shows that a quadratic
Weyl theory is less stable than a quadratic Riemann scalar $R^2$. In particular, in the
very interesting article Ref. \refcite{barrow-hervik}, Barrow and Hervik addressed the
anisotropic cases of Bianchi $I$ and
$II$. The most interesting results of Barrow and Hervik are the exact solutions for
quadratic theories of the same type investigated in this present article.
Instead of the metric, the field equations in Ref. \refcite{barrow-hervik} are written
in a different set of variables which by now is a well known procedure used in the context
of dynamical systems in cosmology.\cite{dsc}. Homogenous solutions in the context of quadratic
gravity was also addressed by Ref. \refcite{resto}.
There is an interesting article by Saridakis, Ref. \refcite{saridakis} in which the
Kantowski Sachs anisotropic case is addressed for quadratic gravity. Schmidt does a review
of higher order gravity theories in connection to cosmology.\cite{hjs}

Also in the context of quadratic theories we have the alternative of
Gauss-Bonnet type $F(G)$ by Odintsov, Nojiri and collaborators Ref. \refcite{odintsov06}
(for recent reviews see Ref. \refcite{odintsov}). When the action depends on a arbitrary
function of the Gauss-Bonnet term, it is not a top invariant and a consistent dynamic
follows from it. Theories of $R^2$ type are also investigated, for example in
 Ref. \refcite{shapiro-odintsov}.

In our case the Gauss-Bonnet term is understood as a surface term and discarded.

We have previously addressed the Bianchi $I$ solutions in
Refs. \refcite{sandro,daniel-sandro}. Stability is also discussed there.
 More recently we have submitted two articles, one in which we investigate how the
 approach to Minkowski space occurs in the Bianchi $I$ case. \cite{daniel-marcio-jcarlos}
 In the other article we present the Bianchi $VII_A$ solutions Ref. \refcite{juliano-daniel}.
 In Ref. \refcite{daniel-marcio-jcarlos} it is obtained that the solution is a superposition
 of a pure tensorial component, and a pure scalar component. Thus the approach to Minkowski
 space involves the production of scalar and tensorial gravitational waves. Speculations
 about the validity of the semi-classical argument can be raised, anyway, in a certain
 sense the quadratic counterterms which we consider are the most natural ones expected
 from the renormalization of a quantum field.\cite{christensen}

The purpose of this article is to cast the dynamical equations of motion in a 3+1
decomposition for the zero vorticity case. A time like and geodesic vector can be
defined and we impose the homogeneity on the 3-space. Apparently it is the first time
the equations are presented in this form. The Bianchi $I$ case is revisited but now the
equations are written in an analytical fashion. We intend to use the expressions obtained
in this work in our future works. Most of the article is there just for completeness.

The article is organized as follows. In section 2 a brief exposition of the vacuum
polarization by the external gravitational field. In section 3 we present the $3+1$
decomposition for any Bianchi type. In the section 4 we specialize to the Bianchi I case.
 And the conclusions are presented in section 5.

\section{The Divergent Counterterms}
Since the contribution of other spin fields to the effective action, are of the same
type for the scalar field, see for example Ref. \refcite{christensen}, we will considered
a quantum scalar field in a curved classical background only
\[
S=\frac{1}{2}\int d^4x\sqrt{-g}\left( \partial_k\phi\partial^k\phi -m^2\phi^2-\xi R\phi^2\right).
\]
After an integration in parts in the above action, the gaussian integral results as
\begin{eqnarray*}
&&e^{-iW}=\int D\phi \exp\left\{\frac{-i}{2}\int d^4x\sqrt{-g}\phi
\left[ \nabla_a\nabla^a+m^2+\xi R\right]\phi\right\}\\
&&e^{-iW}=[\det (iF/2\pi M^2)]^{-1/2}\\
&&W=\frac{i}{2}\mbox{ln} (\mbox{det}(iF/2\pi M^2))=\frac{i}{2}\mbox{tr } \mbox{ln} (F) + C,
\end{eqnarray*}
where $\langle x|F| y\rangle=(\nabla_a\nabla^a+m^2+\xi R)\delta(x-y)/\sqrt{-g}$, $M$ and
$C$ are constants, and the trace means
$\mbox{tr}B=\lim_{x\rightarrow y}\int d^4x\sqrt{-g}\langle x|B| y \rangle$.
The effective Lagrange function is obtained as a Mellin transform of the Schr\"odinger kernel
\begin{eqnarray*}
&&\delta W=\frac{i}{2}\mbox{tr } \left( \frac{\delta F}{F}\right)=\frac{i}{2}\mbox{tr}
\left\{ i\int _ 0^\infty dse^{-isF}\delta F\right\}\\
&&\delta W=- \frac{i}{2}\delta \mbox{tr} \left\{ \int _ 0^\infty ds\frac{e^{-isF}}{s}\right\}\\
&&W=- \frac{i}{2}\int d^4x\sqrt{-g}  \int _ 0^\infty ds\frac{K(s,x)}{s},
\end{eqnarray*}
where the kernel $K(s,x)=\lim_{x\rightarrow x^\prime}K(s,x,x^\prime)$, satisfies a
Schr\"odinger type equation
\[
\square K(s,x,x^\prime) +\xi RK(s,x,x^\prime)+m^2K(s,x,x^\prime)=
i \frac{\partial}{\partial s} K(s,x,x^\prime),
\]
together with the boundary condition
$\lim_{s\rightarrow 0}K(s,x,x^\prime)=\delta(x-x^\prime)$.
This equation can be solved perturbatively,
\[
K(s,x)=\lim_{x\rightarrow x^\prime}K(s,x,x^\prime)=
-\frac{i}{(4\pi s)^2}e^{-i(\sigma/2s+m^2s)}\sum_{n=0}^\infty a_n(is)^n
\]
where $\sigma$ is half of the length squared of the geodesic
\[
\sigma=\frac{1}{2}\int_0^\lambda d\lambda^\prime
\left( g_{ab}\frac{dx^a}{d\lambda^\prime}\frac{dx^b}{d\lambda^\prime}\right) ^{1/2}
\]
and the $a_n$ are known as Seeley-de Witt coefficients. The ones connected
to divergencies are just the first three
\begin{eqnarray*}
&&a_0=1\\
&&a_1=\left( \frac{1}{6}-\xi \right)R\\
&&a_2=\frac{1}{180}\left( R^{abcd}R_{abcd}-R^{ab}R_{ab}\right)-
\frac{1}{6}\left( \frac{1}{5}-\xi \right)\nabla^a\nabla_a R+\frac{1}{2}
\left(\frac{1}{6} -\xi \right)R^2.
\end{eqnarray*}
\[
\mathcal{L}_{div}=\frac{\sqrt{-g}}{32\pi^2}
\int_0^\infty\frac{d\tau}{\tau^3}e^{-m^2\tau}\sum_{n=0}^2 a_n\tau^n.
\]
The term $\nabla_a\nabla^a R$ is a total derivative and will be omitted.
Also making use of the Gauss-Bonnet surface term
\[
\int d^4x\sqrt{-g}(R^{abcd}R_{abcd}-4R^{ab}R_{ab}+R^2),
\]
the square of the Riemann tensor term can be written in a
convenient combination $R_{ab}R^{ab} -1/3R^2$ which is dynamically equivalent
to a term proportional to the square of the Weyl tensor $C_{abcd}C^{abcd}$.
 A theory without these specific counterterms is certainly inconsistent,
 except for very particular situations.

With this in mind, the field equations for the semiclassical theory, are obtained performing
metric variations in the gravitational Lagrangian
\begin{equation}
\mathcal{L}=\sqrt{-g}\left[-\Lambda+R+
\alpha\left(R_{ab}R^{ab}-\frac{1}{3}R^{2}\right)+\beta R^{2}\right],
\label{acao}
\end{equation}
where $\alpha$
and $\beta$ are constants. Classical sources are not taken into account in this present work.
For the spatially homogenous space they are described
by the tensor $E=E_{ab}\omega^{a}\otimes\omega^{b}$, $\omega^0=dt$,
\begin{equation}
E_{ab}\equiv G_{ab}+\frac{1}{2}g_{ab}\Lambda-\left(\beta-\frac{1}{3}\alpha\right
)H_{\: ab}^{(1)}-\alpha H_{\: ab}^{(2)}=0,\label{eq.campo}
\end{equation}
where
\begin{eqnarray*}
&&G_{ab}=R_{ab}-\frac{1}{2}g_{ab}R,\\
&&H_{ab}^{(1)}=\frac{1}{2}g_{ab}R^{2}-2RR_{ab}-2g_{ab}\square R+2R_{;ab},\\
&&H_{ab}^{(2)}=\frac{1}{2}g_{ab}R^{cd}R_{cd}-\square R_{ab}-
\frac{1}{2}g_{ab}\square R+R_{;ab}-2R^{cd}R_{cbda}.
\end{eqnarray*}
First let us emphasize that every Einstein space satisfying $R_{ab}=
g_{ab}\Lambda/2$ is an exact solution of \eqref{eq.campo},
apparently first realized in Ref. \refcite{barrow-hervik}. As any other
metric theory, the covariant divergence of the above tensors are zero. And the the
zero divergence $\nabla_aE^a_b=0$ implies that $E_{00}$ and $E_{0\alpha}=0$, are in
fact constraints, while the dynamical equations are contained in the spatial part
$E_{\alpha\beta}=0$, see for example Ref. \refcite{stephani}, p. 165.
\section{3+1 Decomposition}
We briefly develop the general homogenous case, for more details see
 Ref. \refcite{smk}. The right-invariant basis are the Killing vectors, and satisfy
 the Lie algebra
\[
[\xi_\alpha,\xi_\beta]=-C^\nu_{\alpha\beta}\xi_\nu,
\]
where the greek indices run from $1$ to $3$. On the other hand the left-invariant
basis vectors satisfy
\begin{eqnarray*}
&&[e_\alpha,e_\beta]=C^\nu_{\alpha\beta}e_\nu \\
&&[e_0,e_\alpha]=0.
\end{eqnarray*}
The connection is defined as
\[ \nabla_a e_b=\Gamma^c_{ba}e_c.
\]
The anti symmetric part of the connection follows from
\[
\nabla_\beta e_\alpha-\nabla_\alpha e_\beta=[e_\beta,e_\alpha],
\]
which can be regarded as the definition of zero torsion. Metricity $\nabla_cg_{ab}=0$
implies that
\[
\Gamma_{abc}=\frac{1}{2}\left( g_{ab|c}+g_{ac|b}-g_{bc|a}\right)+
\frac{1}{2}\left(-C_{abc}+C_{bac}+C_{cab}\right),
\]
where the $|$ means the directional derivative $f_{|a}=e_a^i\partial_i f$.
We restrict the analysis to rotation free models in this present work
\[
ds^2=-dt^2+h_{\alpha\beta}\omega^\alpha\otimes \omega^\beta,
\]
where $\omega^\alpha$ is the left invariant $1-$ form basis, dual to $e_\beta$.
Once the structure constants are known, in fact, the $1-$ form basis can be found by
solving Cartan first structure equation
\[
d\omega^\alpha=-\frac{1}{2}C^\alpha_{\mu\nu}\omega^\mu\wedge\omega^\nu.
\]
The time like vector field $u^a=(1,0,0,0)$ is geodesic and rotation free, as can be
easily checked for the above line element and connection. In this particular case the
extrinsic curvature
\begin{eqnarray*}
&&K_{\alpha\beta}=\nabla_\alpha u_{\beta}=\Gamma^0_{\alpha\beta}=\frac{1}{2}\dot{h}_{\alpha\beta}\\
&&K_\alpha^\beta=h^{\beta\mu}K_{\mu\alpha}=\Gamma^\beta_{\alpha 0}=\Gamma^\beta_{0\alpha},
\end{eqnarray*}
and the spatial part of the connection reads
\[
\Gamma_{\alpha\beta\gamma}=\frac{1}{2}\left( -C_{\alpha\beta\gamma}+
C_{\beta\alpha\gamma}+C_{\gamma\alpha\beta}\right).
\]
The $3$ and $4$ Riemann curvature follows
\begin{eqnarray*}
&&{}^3R^\alpha_{\beta\mu\nu}=\Gamma^\alpha_{\rho \mu}\Gamma^\rho_{\beta\nu}-
\Gamma^\alpha_{\rho \nu}\Gamma^\rho_{\beta\mu}-C^\rho_{\mu\nu}\Gamma^\alpha_{\beta\rho} \\
&&R^\alpha_{\beta\mu\nu}= {}^3R^\alpha_{\beta\mu\nu}+K^\alpha_\mu K_{\beta\nu}-
K^\alpha_ \nu K_{\beta\mu}\\
&&R^0_{\alpha\beta\mu}=-C^\rho_{\beta\mu}K_{\alpha\rho}+
K_{\rho\beta}\Gamma^\rho_{\alpha\mu}-K_{\rho \mu}\Gamma^\rho_{\alpha\beta}\\
&&R^0_{\alpha0\beta}=\dot{K}_{\alpha\beta}-K_{\rho\beta}K^\rho_\alpha,
\end{eqnarray*}
note that $\dot{K}=\partial_t\left( K^1_1+K^2_2+K^3_3\right)$.
The Ricci tensor and Riemann scalar follow from contractions of the above tensors
\begin{eqnarray*}
&&R_{\alpha\beta}={}^3R_{\alpha\beta}+KK_{\alpha\beta}-
2K_\alpha^\rho K_{\rho\beta}+\dot{K}_{\alpha\beta}\\
&&R_{0\alpha}=C^\rho_{\alpha\beta}K^\beta_\rho-K^\rho_\alpha C^\beta_{\rho\beta}\\
&&R_{00}=-\dot{K}-K_{ij}K^{ij}\\
&&R={}^3R+2\dot{K}+K_{ij}K^{ij}+K^2.
\end{eqnarray*}
This theory depends on the derivatives of the Ricci tensor and Riemann scalar
\begin{eqnarray*}
&&\nabla_a\nabla_0R=\delta_{a 0}\ddot{R} \\
&&\nabla_\alpha\nabla_\beta R =-K_{\alpha\beta}\dot{R}\\
&&\square R=-\ddot{R}-K\dot{R}\\
&&\square R_{00}=-\ddot{R}_{00}-\dot{R}_{00}K+
2R_{\alpha\beta}K^\alpha_\rho K^{\rho\beta}+2K_\rho^\alpha C^\rho_{\beta\alpha}R_0^\beta +
2C^\rho_{\alpha\rho}K^{\alpha\beta}R_{0\beta}\\
&&+2K_{\alpha\beta}K^{\alpha\beta}R_{00} \\
&&\square R_{0\alpha}=-\ddot{R}_{0\alpha}+2\dot{R}_{0\rho}K^\rho_\alpha-K\dot{R}_{0\alpha}+
2K^{\rho \beta}\Gamma^\nu_{\alpha\beta}R_{\nu\rho} +
R_{00}\left( -C^\rho_{\alpha\nu} K^\nu_\rho+C^\rho_{\nu\rho}K^\nu_\alpha\right) \\
&&+ R_{0\alpha}K^{\mu\nu}K_{\mu\nu}+R_{0\nu}\left( \dot{K}^\nu_\alpha +KK_\alpha^\nu+
\Gamma^\mu_{\alpha\beta}\Gamma^\nu_{\mu\gamma}h^{\beta\gamma}+
C^\gamma_{\mu\gamma}\Gamma^\nu_{\alpha\rho}h^{\rho\mu}\right) +
2R_{0\nu}K^{\nu \mu}K_{\mu\alpha}\\
&& +R_\alpha^\nu\left( K^\rho_\mu C^\mu_{\nu \rho}+C^\mu_{\rho\mu}K_\nu^\rho\right) \\
&&\square R_{\alpha\beta}=-\frac{1}{2}\ddot{R}_{\alpha\beta}+
2\dot{R}_{\rho\beta}K^\rho_\alpha-\frac{1}{2}K\dot{R}_{\alpha\beta}+
R_{\nu\rho}\Gamma^\rho_{\alpha\mu}\Gamma^\nu_{\beta\gamma}h^{\mu\gamma}+
R_{00}K_\alpha^\mu K_{\mu\beta}\\
&&-R_{\mu\nu}K^\mu_\alpha K^\nu_\beta  + R_{\beta\nu}\left( \dot{K}^\nu_\alpha +KK_\alpha^\nu+
 \Gamma^\rho_{\alpha \mu}\Gamma^\nu_{\rho \gamma}h^{\mu \gamma}+
 C^\mu_{\rho\mu}\Gamma^\nu_{\alpha\theta}h^{\theta\rho}\right) \\
 && +R_{\alpha0}\left( C^\rho_{\beta\nu}K^\nu_\rho+C^\mu_{\rho\mu}K^\rho_\beta\right)+\alpha \leftrightarrow \beta,
\end{eqnarray*}
after the substitution of the above expressions into \eqref{eq.campo}, it can be seen
that the higher time derivatives are cointained in the spatial parts of $H^{(1)}$ and $H^{(2)}$.

\section{Bianchi $I$}
We shall apply the preceding expressions to the particular spatially flat case.
The reason is that the
${}^3 R^\alpha_{\beta\gamma\delta}$, acts as a generalized potential much in
the same sense as in the mixmaster case. In this sense, ${}^3R^\alpha_{\beta\gamma\delta}$,
can decrease and become irrelevant and the dynamics can be arbitrarily  close to the
Bianchi $I$ case.

This is the abelian case in which all the structure constants are zero.
Of course ${}^3R_{\alpha\beta\mu\nu}=0$. We will restrict ourselves to diagonal
extrinsic curvature $K_{\alpha\beta}=\mbox{diag }[K_{1},K_{2},K_{3}]$, which is consistent
with a diagonal metric also $h_{\alpha\beta}=\mbox{diag }[a_1^2,a_2^2,a_3^2]$.
Note that $\dot{K}_{\alpha\mu}h^{\mu\beta}=\delta^\beta_\alpha( \dot{H}_\beta+2(H_\beta)^2)$, where $H_\beta=\dot{a}_\beta/a_\beta$
are the Hubble constants in each direction, and as usual, the trace $K=H_1+H_2+H_3=\theta$,
is the expansion
\begin{eqnarray*}
&&R_\alpha^\beta=\delta_\alpha^\beta \left( \theta H_\beta +\dot{H}_\beta \right)\\
&&R_{0\alpha}=0\\
&&R_{00}=-\dot{\theta}-H_\alpha H_\beta\delta^{\alpha\beta}\\
&&R=2\dot{\theta}+\theta^2+H_\alpha H_\beta\delta^{\alpha\beta}.
\end{eqnarray*}
In the following we quote the result for the contribution coming from the $R^2$ term, 
\begin{eqnarray*}
&&H^{(1)}_{00}=-\frac{1}{2}\left( 2\dot{\theta}+\theta^2+
H_\mu H_\nu\delta^{\mu\nu}\right)^2+2\left(2\dot{\theta}+
\theta^2+H_\mu H_\nu\delta^{\mu\nu}\right)(\dot{\theta}+
H_\mu H_\nu\delta^{\mu\nu})\\
&&-4\theta(\ddot{\theta}+\theta\dot{\theta}+\dot{H}_\mu H_\nu\delta^{\mu\nu})\\
&&H^{(1)}_{0\alpha}=0\\
&&H^{(1)\alpha}_{\;\;\;\;\beta}=\delta^\alpha _\beta\left[ \frac{1}{2}
\left( 2\dot{\theta}+\theta^2+H_\mu H_\nu\delta^{\mu\nu} \right)^2 -
2\left( 2\dot{\theta}+\theta^2+H_\mu H_\nu\delta^{\mu\nu} \right)
\left( \theta H_\alpha +\dot{H}_\alpha \right)\right. \\
&&\left. +4\left(\theta-H_\alpha\right)\left( \ddot{\theta}+
\dot{\theta}\theta+\dot{H}_\mu H_\nu\delta^{\mu\nu}\right)
+4\left( \dddot{\theta} +\dot{\theta}^2+\theta\ddot{\theta}+
\ddot{H}_\mu H_\nu\delta^{\mu\nu}
+ \dot{H}_\mu\dot{H}_\nu\delta^{\mu\nu}\right) \right], 
\end{eqnarray*}
the spatial trace of which, $H^{(1)\alpha}_{\;\;\;\;\alpha}=H^{(1)1}_{\;\;\;\; 1}+H^{(1)2}_{\;\;\;\; 2}+H^{(1)3}_{\;\;\;\; 3}$, 
\begin{eqnarray*}
&&H^{(1)\alpha}_{\;\;\;\;\alpha}=\frac{3}{2}\left( 2\dot{\theta}+\theta^2+H_\mu H_\nu \delta^{\mu\nu}\right)^2
+12\dddot{\theta}+8\dot{\theta}^2+20\theta\ddot{\theta}+2\theta^2\dot{\theta}-2\theta^4\\
&&-2H_\mu H_\nu\delta^{\mu\nu}\left(\theta^2+\dot{\theta}\right)+8\theta\dot{H}_\mu H_\nu\delta^{\mu\nu}
+12\ddot{H}_\mu H_\nu\delta^{\mu\nu}+12\dot{H}_\mu \dot{H}_\nu \delta^{\mu\nu}.
\end{eqnarray*}
And the contribution from the $R_{ab}R^{ab}$ term
\begin{eqnarray*}
&&H^{(2)}_{00}=-\frac{1}{2}\left[ \left(\dot{\theta} + H_\mu H_\nu\delta^{\mu\nu}\right)^2+
\sum_\nu\left( \theta H_\nu +\dot{H}_\nu\right)^2 \right] +\dot{\theta}^2 
-\ddot{H}_\mu H_\nu\delta^{\mu\nu} \\
&&-\dot{H}_\mu\dot{H}_\nu\delta^{\mu\nu} +2H_\mu H_\nu\delta^{\mu\nu}
\left( \dot{\theta} +H_\rho H_\tau\delta^{\rho\tau}\right)+
2\dot{H}_\mu \left( \theta H_\nu+
\dot{H}_\nu\right)\delta^{\mu\nu}\\
&&-\theta\left( \ddot{\theta} +\theta \dot{\theta} +
3\dot{H}_\mu H_\nu\delta^{\mu\nu}\right)\\
&&H^{(2)}_{0\alpha}=0\\
&&H^{(2)\alpha}_{\;\;\;\;\beta}=\delta^\alpha_\beta\left\{
\frac{1}{2}\left[ \left(\dot{\theta} + H_\mu H_\nu\delta^{\mu\nu}\right)^2+
\sum_\nu\left( \theta H_\nu +\dot{H}_\nu\right)^2 \right] \right.\\
&&\left.+2(H_\alpha)^2\left(\dot{\theta}+H_\mu H_\nu\delta^{\mu\nu}\right)+ \left( \ddot{\theta} H_\alpha+2 \dot{\theta} \dot{H}_\alpha+
\theta \ddot{H}_\alpha +\dddot{H}_\alpha\right)\right.\\
&&\left.-2H_\alpha\left( \ddot{\theta}+\theta\dot{\theta}+
\dot{H}_\mu H_\nu\delta^{\mu\nu} \right) + \left[
\left( \dddot{\theta}+\dot{\theta}^2+\theta\ddot{\theta}+
\ddot{H}_\mu H_\nu\delta^{\mu\nu}+\dot{H}_\mu \dot{H}_\nu\delta^{\mu\nu}\right)\right.\right.\\
&&\left.\left.+ \theta\left( \ddot{\theta}+
\theta\dot{\theta}+\dot{H}_\mu H_\nu\delta^{\mu\nu} \right) \right]  -2\left[
H_\alpha\delta^{\mu\nu}H_\nu\left( \theta H_\mu+\dot{H}_\mu\right)\right.\right.\\
&&\left.\left.+\left(\dot{\theta}+H_\mu H_\nu\delta^{\mu\nu}\right)
\left( \dot{H}_\alpha+(H_\alpha)^2\right)\right]+\theta\left(\dot{\theta}H_\alpha+\theta\dot{H}_\alpha+\ddot{H}_\alpha\right) \right\},
\end{eqnarray*}
with spatial trace  
\begin{eqnarray*}
&&H^{(2)\alpha}_{\;\;\;\;\alpha}=\frac{3}{2}\left[(\dot{\theta}+H_\mu H_\nu\delta^{\mu\nu})^2+\sum_\nu\left( \theta H_\nu +\dot{H}_\nu\right)^2 \right]+7\theta\ddot{\theta} + 3\theta^2\dot{\theta} \\
&&+3\dot{\theta}^2+4\dddot{\theta}+3\ddot{H}_\mu H_\nu\delta^{\mu\nu}-\theta\dot{H}_\mu H_\nu\delta^{\mu\nu}-2H_\mu H_\nu \delta^{\mu\nu}\left( \dot{\theta}+\theta^2\right)+3\dot{H}_\mu\dot{H}_nu\delta^{\mu\nu}.
\end{eqnarray*}
Now we can use the sum of the spatial part of field equations \eqref{eq.campo}
to obtain the higher derivative of the expansion $\theta$, giving the equivalent of Raychaudhuri's equation
\begin{eqnarray*}
&&\dddot{\theta}=\frac{1}{12\beta}\left\{-2\dot{\theta}-
\frac{1}{2}\theta^2-\frac{3}{2}H_\mu H_\nu\delta^{\mu\nu} +\frac{3}{2}\Lambda\right.\\
&&\left.-\left( \beta-\frac{\alpha}{3}\right)\left[
\frac{3}{2}\left( 2\dot{\theta}+\theta^2+H_\mu H_\nu \delta^{\mu\nu}\right)^2
+8\dot{\theta}^2+20\theta\ddot{\theta}+2\theta^2\dot{\theta}-2\theta^4\right.\right.\\
&&\left.\left.-2H_\mu H_\nu\delta^{\mu\nu}\left(\theta^2+\dot{\theta}\right)+8\theta\dot{H}_\mu H_\nu\delta^{\mu\nu}
+12\ddot{H}_\mu H_\nu\delta^{\mu\nu}+12\dot{H}_\mu \dot{H}_\nu \delta^{\mu\nu}
 \right] \right.\\
 &&\left.-\alpha\left[
 \frac{3}{2}\left((\dot{\theta}+H_\mu H_\nu\delta^{\mu\nu})^2+\sum_\nu\left( \theta H_\nu +\dot{H}_\nu\right)^2 \right)+7\theta\ddot{\theta} + 3\theta^2\dot{\theta}\right.\right. \\
&&\left.\left.+3\dot{\theta}^2+3\ddot{H}_\mu H_\nu\delta^{\mu\nu}-\theta\dot{H}_\mu H_\nu\delta^{\mu\nu}-2H_\mu H_\nu \delta^{\mu\nu}\left( \dot{\theta}+\theta^2\right)+3\dot{H}_\mu\dot{H}_nu\delta^{\mu\nu}
 \right]\right\}.\\
\end{eqnarray*}
Again, using the field equations \eqref{eq.campo} the higher derivatives
$\dddot{H}_\alpha$ can be isolated, and $\dddot{\theta}$ can be substituted in the
following expression yielding a consistent dynamical system
\begin{eqnarray*}
&&\dddot{H}_\alpha=\frac{1}{\alpha}\left\{ \theta H_\alpha+\dot{H}_\alpha-\dot{\theta}-
\frac{\theta^2}{2}-\frac{1}{2}H_\mu H_\nu\delta^{\mu\nu}+\frac{1}{2}\Lambda\right.\\
&&\left. -\left(\beta-\frac{1}{3}\alpha \right)\left[ \frac{1}{2}\left( 2\dot{\theta}+
\theta^2+H_\mu H_\nu\delta^{\mu\nu} \right)^2 -2\left( 2\dot{\theta}+
\theta^2+H_\mu H_\nu\delta^{\mu\nu} \right)
\left( \theta H_\alpha +\dot{H}_\alpha \right)\right. \right.\\
&&\left. \left.+4\left(\theta-H_\alpha\right)\left( \ddot{\theta}+
\dot{\theta}\theta+\dot{H}_\mu H_\nu\delta^{\mu\nu}\right)
+4\left( \dddot{\theta} +\dot{\theta}^2+\theta\ddot{\theta}+
\ddot{H}_\mu H_\nu\delta^{\mu\nu}
+ 2\dot{H}_\mu\dot{H}_\nu\delta^{\mu\nu}\right) \right]\right.\\
&&\left. -\alpha\left[
\frac{1}{2}\left[ \left(\dot{\theta} + H_\mu H_\nu\delta^{\mu\nu}\right)^2+
\sum_\nu\left( \theta H_\nu +\dot{H}_\nu\right)^2 \right] +
2(H_\alpha)^2\left(\dot{\theta}+H_\mu H_\nu\delta^{\mu\nu}\right)\right.\right.\\
&&\left.\left.+ \left( \ddot{\theta} H_\alpha+2 \dot{\theta} \dot{H}_\alpha+
 \theta \ddot{H}_\alpha \right)-2H_\alpha\left( \ddot{\theta}+\theta\dot{\theta}+
 \dot{H}_\mu H_\nu\delta^{\mu\nu} \right) + \left[ \theta\left( \ddot{\theta}+\theta\dot{\theta}+
\dot{H}_\mu H_\nu\delta^{\mu\nu} \right)\right.\right.\right.\\
&&\left.\left.\left.+\left( \dddot{\theta}+\dot{\theta}^2+
\theta\ddot{\theta}+\ddot{H}_\mu H_\nu\delta^{\mu\nu}+
\dot{H}_\mu \dot{H}_\nu\delta^{\mu\nu}\right) \right] -2\left[\left(\dot{\theta}+
H_\mu H_\nu\delta^{\mu\nu}\right)\left( \dot{H}_\alpha+
(H_\alpha)^2\right)\right.\right.\right.\\
&&\left.\left.\left.+H_\alpha\delta^{\mu\nu}H_\nu\left( \theta H_\mu+\dot{H}_\mu\right)\right]+\theta\left(\dot{\theta}H_\alpha+\theta\dot{H}_\alpha+\ddot{H}_\alpha\right)
\right]
\right\}.
\end{eqnarray*}
In the above expressions only the spatial part of \eqref{eq.campo} is used, while not shown here is the $00$ component of $E_{ab}$ which acts as a hamiltonian constraint. Let us emphasize that $\dddot{\theta}=\dddot{H}_1+\dddot{H}_2+\dddot{H}_3$, thus this last two equations are not independent.

\section{Conclusions}
The semi-classical theory consider the back reaction of quantum fields in a classical geometric
background. It began about forty years ago with De Witt Ref. \refcite{DeWitt}, and since then, its
consequences and applications are still under research, see for example Ref. \refcite{Hu}.

Different from the usual Einstein-Hilbert action, the one loop effective gravitational
action surmounts
to quadratic theories in curvature, see for example Refs. \refcite{DeWitt,liv}. It is the
 gravitational version of the Heisenberg-Euler electromagnetism. As it is well known, vacuum
 polarization introduces non linear corrections into Maxwell electrodynamics,\cite{schwinger}
 first obtained by Heisenberg- Euler.\cite{heisenberg}

In this article we present the decomposition $3+1$ for arbitrary spatially homogenous
space times for this particular effective quadratic theory. It turns out that for this
particular gravitational theory \eqref{acao}, every Einstein space satisfying
$R_{ab}=g_{ab}\Lambda/2$ is an exact solution, this includes of course the vacuum case
$\Lambda=0$. Anyway the most interesting is the BKL oscillatory approach to the singularity
which occurs for vacuum in Einstein's context.\cite{bkl} Since the mixmaster solution is a
vacuum $R_{ab}=0$ solution for the Bianchi $IX$ case it will occur exactly the same
in the effective theory considered in \eqref{acao}. 

In this sense a better understanding of the Bianchi $I$ case for the quadratic theories
seems attractive. The $3$ curvature also can be understood as a potential and while approaching
the singularity its influence in the dynamics can decrease almost to zero. Thus, an arbitrary
Bianchi solution of the quadratic gravity can approach very much the Bianchi $I$ case.
We intend to use this $3+1$ decomposition, numerically, in future works. Particular
Bianchi $I$ analytic exact solutions for a quadratic theory identical to this one were already found
by Ref. \refcite{barrow-hervik}. Also for $f(R)$ gravity models including
the $R+R^2$ case, Bianchi I space-times were studied in Ref. \refcite{gurovich}, 
where it was shown that equations for the anisotropic part of the
metric can be integrated.

We emphasize that the approach of this present work is facing the theory \eqref{acao}  as an effective, and classical theory. While considering it as a candidate for a quantized gravity does introduces ghosts. Which together with tachyons and the additional degrees of freedom in contrast to Einstein's theory, makes it improbable that the BKL solution will be the generic one in this case.

\section*{Acknowledgments}
The author wishes to thank the Friedmann seminar organizing committee and the
Brazilian projects {\it Nova F\'\i sica no Espa\c co} and INCT-A.

\end{document}